\documentstyle[prl,twocolumn,aps,epsfig]{revtex}
\begin{document}

\draft
\preprint{nucl-th/0003???}

\title{Nonlinear QCD Evolution: Saturation without
Unitarization}

\author{Alexander Kovner$^a$ and 
   Urs Achim Wiedemann$^b$}
\address{$^a$Department of Mathematics and Statistics, University of Plymouth,
2 Kirkby Place, Plymouth, PL4 8AA, U.K.\\
$^b$Theory Division, CERN, CH-1211 Geneva 23, Switzerland}

\date{\today}

\maketitle

\begin{abstract}
We consider the perturbative description of saturation based on the 
nonlinear QCD evolution equation of Balitsky and Kovchegov (BK).  
Although the nonlinear corrections lead to saturation of 
the scattering amplitude locally in impact parameter space, 
we show that they do not unitarize the total cross section. 
The total cross section for the scattering of a strongly interacting 
probe on a hadronic target is found to grow exponentially with 
rapidity $t = ln(s/s_0)$, 
$\sigma\propto \exp\{{\alpha_s N_c\over{2\pi}}\epsilon t\}$ 
where $\epsilon$ is a number of order unity. The origin of this 
violation of unitarity is the presence of long range Coulomb fields 
away from the saturation region. The growth of these fields with
rapidity is not tempered by the nonlinearity of the BK equation.
\end{abstract} 
%\pacs{PACS numbers: 25.75.+r, 07.60.ly, 52.60.+h}

\vskip 0.3cm

%%%%%%%%%%%%%%%%%%%%%%%%%%%%%%%%%%%%%%%%%%%%%%%%%%%%%%%%%%%%%%%%%%%%%%

Understanding the growth of total scattering cross sections with 
energy $\sqrt{s}$ is a longstanding problem. The unitarity, 
or Froissart bound states that the total inelastic cross section for 
the scattering of a hadronic projectile on a hadronic target can not 
grow faster than 
\begin{equation}
  \sigma<\pi d^2\ln^2(s/s_0)\, ,
  \label{unitarity}
\end{equation}
where $d$ is some typical hadronic scale and $t = ln(s/s_0)$ the
rapidity. While QCD, the theory of hadronic interactions, is a 
unitary theory and
therefore satisfies this unitarity bound, there is no
guarantee that perturbative calculations preserve this property. 
In fact, the linear perturbative evolution equation due to
Balitsky, Fadin, Kuraeev and Lipatov (BFKL) implies an exponential
growth of $\sigma$ with $t$, thus violating unitarity.

Following the pioneering works of \cite{glr,mcraju}, there has been
recent progress in high energy hadronic scattering in a derivation 
of a nonlinear evolution 
equation~\cite{balitsky,kovchegov,weigert} which tames the
BFKL-type growth. These equations resum the nonlinear corrections to 
the QCD evolution with rapidity (energy) to all orders in partonic 
density and to first order 
in the QCD coupling. While previous studies of these equations
assumed translational invariance in the impact parameter plane,
here we explore for the first time their impact parameter 
dependence. For the equations first derived by 
Balitsky \cite{balitsky}, we show that the total cross section
does not unitarize but grows exponentially with $t$.

The first in the hierarchy of nonlinear BK evolution equations
which govern the evolution of correlation functions of the gluon 
fields in the target with $t$, reads\cite{balitsky} 
\begin{eqnarray}
&&{d\over d t}{\rm Tr} \langle 1-U^\dagger(x)U(y)\rangle
   ={\alpha_s\over2\pi^2} \int d^2z\, {(x-y)^2\over(x-z)^2(y-z)^2}
   \nonumber\\
&& \langle N_c{\rm Tr}[U^\dagger(x)U(y)]-{\rm Tr}[U^\dagger(x)U(z)]{\rm
   Tr}[U^\dagger(z)U(y)]\rangle\, , 
   \label{bal}
\end{eqnarray}
where $U(x)$ is the eikonal scattering amplitude of a fundamental
probe on the target characterized by some distribution
of gluon fields $A_\mu$. In the gauge used in~\cite{balitsky}
\begin{equation}
  U(x)=P\exp\Big [i\int dx^-T^aA_a^+(x)\Big ]\, . 
  \label{amp}
\end{equation}
The averaging in (\ref{bal}) is taken over the ensemble characterizing
the target. In the large $N_c$ limit, (\ref{bal}) simplifies to a closed
equation for the scattering probability $N(x,y)$ of 
a colour singlet dipole with charges at points $x,y$ 
\begin{equation}
  N(x,y)={1\over N_c}{\rm Tr}\langle 1-U^\dagger(x)U(y)\rangle\, .
\end{equation}
This equation was independently derived by 
Kovchegov~\cite{kovchegov} 
in the colour dipole approach of~\cite{muellerdip}
\begin{eqnarray}
  &&{d\over dt}N(x,y) = {\alpha_sN_c\over 2\pi^2}\int
                        d^2z{(x-y)^2\over(x-z)^2(y-z)^2}
  \nonumber \\
&&\qquad  [N(x,z)+N(y,z)-N(x,y)-N(x,z)N(z,y)] \, .
  \label{kov}
\end{eqnarray}
Weigert~\cite{weigert} suceeded to reformulate Balitsky's hierarchy
in terms of a nonlinear stochastic process,
\begin{eqnarray}
  {dU(x)\over dt} &=& g U(x)iT^a \hspace{-.2cm}
     \int \hspace{-.2cm}{d^2z\over \sqrt{4\pi^3}} {(x-z)_i\over (x-z)^2} 
     \hspace{-.1cm}
     \left[1-\tilde U^\dagger(x)\tilde
              U(z)\right]^{ab}\hspace{-.3cm} \xi^b_i(z)
     \nonumber \\
     && - \frac{i\alpha_s}{2\pi^2}  
     \int d^2z {1\over (x-z)^2}\, 
     {\rm Tr}[T^a \tilde U^\dagger(x)\tilde U(z)]\, .
     \label{wei}
\end{eqnarray}
Here $U(x)$ and $\tilde U(x)$ are the unitary matrices
(\ref{amp}) in the fundamental and adjoint representations,
respectively. The noise is characterised by Gaussian local
correlations
\begin{equation}
  \langle \xi^a_i(t',z')\xi^b_j(t'',z'')\rangle 
  = \delta^{ab}\delta_{ij}\delta(t'-t'') \delta(z'-z'')\, .
\end{equation}
This simple Langevin equation gives rise to an infinite
number of equations for correlators of $U$ which coincide with
those derived in \cite{balitsky}. 

Nonlinear evolution equations were also derived in 
the Wilson renormalization group approach~\cite{jklw}. 
While results from this approach coincide with 
(\ref{bal},\ref{kov}) in a certain limit, the question
whether they are generally equivalent or incorporate
different physics is still open. Here, we
focus entirely on the BK equations (\ref{bal},\ref{kov},\ref{wei}).

From the first numerical~\cite{lt,braun,gms} and 
analytical~\cite{kov1,el} studies of the BK eqs. (\ref{bal},\ref{kov}) 
the following consistent picture emerges: Suppose one starts the
evolution from the initial condition of small target fields (or
$N(x,y)\ll 1$ for all $x,y$). Then initially the evolution follows
the BFKL equation, since the nonlinear term in (\ref{kov}) is
negligible. As the scattering probability approaches unity, the
nonlinear term kicks in and eventually the growth stops as the RHS
of (\ref{kov}) vanishes for $N(x,y)=1$. The larger dipoles
[$(x-y)^2 \gg 1/Q^2_s(t)$] saturate earlier, the smaller dipoles
follow at later "time" $t$. These features are contained in the
simple parametrization~\cite{gw} 
\begin{equation}
  N(x,y)=1-\exp\hspace{-.1cm}\Big [ -(x-y)^2Q^2_s(t)\Big ]\, .
  \label{golec}
\end{equation}
The saturation momentum $Q_s(t)$ is a growing function of rapidity. 
Its exact dependence on rapidity is not known, but both,
the numerical results \cite{lt} and simple theoretical estimates
\cite{mueller,el} are consistent with the exponential
growth of the form
\begin{equation}
  Q_s(t)=\Lambda\exp\hspace{-.1cm}\Big [ \alpha_s c t\Big ]
  \label{qexp}
\end{equation}
with $c$ of order unity. This physical picture has been
anticipated several years ago in \cite{mueller}.

While the BFKL equation leads to an unphysical exponential 
growth of the scattering probability $N(x,y)$ with $t$,
the nonlinearities of eqs. (\ref{bal},\ref{kov}) tame this
growth such that $N(x,y) < 1$, as required for a probability. 
This is commonly refered to as ``unitarization''. 
However, the ``saturation'' of the scattering probability at fixed 
impact parameter does not insure that the total 
scattering cross section is unitary. We hence refer to the above
phenomenon more accurately as "saturation".

To calculate the total inelastic cross section one has to integrate 
the scattering probability over the impact parameter. Thus in the 
saturation regime
\begin{equation}
\sigma=\pi R^2(t)\, ,
\end{equation}
where $R(t)$ is the size of the region in the transverse plain
for which the scattering probability for hadronic size "dipoles" is
unity. % [Here, the word "dipole" refers not only to a quark - 
% antiquark pair, but to any hadronic state of typical hadronic size.]. 
To satisfy the Froissart bound the radius $R(t)$
should grow at most linearly with $t$.
We now present two simple calculations 
which establish that within the BK evolution the growth of the radius
with rapidity is exponential.

First consider the Langevin equation (\ref{wei}).
Assume that initially, at rapidity $t_0$ the target is black
within some radius $R_0$. This means that for $|z|<R_0$ the matrix
$U(z)$ fluctuates very strongly so that it covers the whole group
space. We concentrate on a point $x$ which is initially outside of
this black region. The matrix $U(x)$ then is close to unity. Thus 
there is no correlation between $U(x)$ and $U(z)$, and the second 
term on the right hand side of (\ref{wei}) can be set to zero. This 
is the random phase approximation introduced in \cite{weigert} and 
used later in \cite{el}.
As the target field ensemble evolves in rapidity, the radius of the
black region grows. As long as the point $x$ stays outside the
black region we can approximate the Langevin equation by
(we drop colour indices which are inessential to our argument)
\begin{equation}
  {d\over dt}U(x)=-\sqrt{{\alpha_s N_c\over{\pi^2}}}\int_{|z|<R}
  d^2z{(x-z)_i\over (x-z)^2}\xi_i(z)\, . 
  \label{langevin}
\end{equation}
This equation neglects contributions to the derivative of $U$ that 
come from gluons originating from the sources outside the black region. 
Those contributions would enhance the growth of $U$, and so by omitting 
them we underestimate the rate of growth of the radius of the black region.
The formal solution of eq. (\ref{langevin}) is
\begin{equation}
  1-U(x,t)= \sqrt{{\alpha_s N_c\over{\pi^2}}}\int_{t_0}^{t}
  \hspace{-.1cm} d\tau \hspace{-.1cm}
  \int_{|z|<R(\tau)} \hspace{-.5cm} d^2z{(x-z)_i\over (x-z)^2}\xi_i(z)\, .
  \label{langevins}
\end{equation}
Squaring it and averaging over the noise term gives
\begin{equation}
  \langle (1-U(x,t))^2\rangle ={\alpha_s N_c\over{\pi^2}}\int_{t_0}^t
  d\tau\int_{|z|<R(\tau)} {d^2z\over (x-z)^2} \, .
  \label{langevinso}
\end{equation}
As long as $x$ is outside the black region % and $|x|>R$ 
we can approximate the integral on the right hand side by
\begin{equation}
  \int_{|z|<R(\tau)}
  d^2z{1\over (x-z)^2}=\pi{R^2(\tau)\over x^2}\, ,
  \label{appr}
\end{equation}
and eq.(\ref{langevinso}) becomes
\begin{equation}
  \langle (1-U(x,t))^2\rangle =
  {\alpha_s N_c\over{\pi}}{1\over x^2} \int_{t_0}^t
  d\tau R^2(\tau) \, .
  \label{langevinsol}
\end{equation}
Now as the black region grows, eventually it reaches the point
$x$. At this rapidity the matrix $U(x)$ will start fluctuating with
the amplitude of order one. Thus when $R(t)=|x|$, the left hand side
of eq. (\ref{langevinsol}) becomes a number of order one, which we call
$1/\epsilon$. We thus have an approximate equation for $R(t)$
\begin{equation}
  {1\over\epsilon}R^2(t)={\alpha_s N_c\over{\pi}} \int_{t_0}^t
  d\tau R^2(\tau)\, .
   \label{radius}
\end{equation}
At large rapidities therefore the radius of the black region
is exponentially large
\begin{equation}
  R(t)=R(t_0)\exp\hspace{-.1cm}\Big [{\alpha_s 
       N_c\over{2\pi}}\epsilon (t-t_0)\Big ]\, .
  \label{radiuss}
\end{equation}
This is our main result. 

We note that while the approximations leading to eq. (\ref{radius})
cease to be valid when the point $x$ is on the boundary of the black 
region, this does not affect our main conclusion.
First, eq. (\ref{appr}) is an underestimate of the integral, 
thus underestimating the growth of $R$.
Second, when $x$ is on the boundary of the black region and 
$z$ in the black region, although the factors $(1-U(x)U^\dagger(z))$ 
and $U(x)$ in eq.(\ref{wei}) are not strictly unity, they are still 
of order one for allmost all points $z$. Thus, although 
we can not determine the exact numerical value of $\epsilon$, the 
functional form of the solution as well as its parametric dependence
is given correctly by eq.(\ref{radiuss}).

Note that eq. (\ref{wei}) refers to the evolution of matrices $U(x)$.
This can be thought of as evolution of the scattering amplitude of a
coloured probe. The preceding derivation thus refers to the growth with
rapidity of the cross section for scattering of a coloured probe.
In a confining theory this is not a physical quantity. However
the physics of BK equation does not incorporate effects
of confinement, and therefore the cross section for a colourless
dipole within the BK framework must grow in the same way.
To establish this point, and to make more explicit the
relation between $\epsilon$ and the BFKL dynamics, we now present
an alternative derivation of eq. (\ref{radiuss}).

To this end we consider the BK evolution as the evolution of the 
projectile \cite{kovchegov}. Suppose at the initial energy the 
projectile is a colour dipole of size $x_0$. It scatters on a 
hadronic target of some size $R_{\rm target}$. 
As is explicit in \cite{kovchegov}, as the energy is increased 
the projectile wave function evolves according to the BFKL equation. 
Thus at rapidity $t$ the density of dipoles of size $x$ at transverse \
distance $r$ from the original dipole is given by the BFKL expression 
(see for example \cite{forshaw}):
\begin{equation}
  n(x_0,x,r,t)={32\over x^2}{\ln {16r^2\over x_0 x}\over(\pi
  a^2t)^{3/2}} \exp\hspace{-.1cm}\Bigg [
  \omega t-\ln {16r^2\over x_0 x} -{\ln^2
  {16r^2\over x_0 x}\over a^2t}\Bigg ]
  \label{density}
\end{equation}
with $\omega=4\ln 2 N_c\alpha_s/\pi$ and
$a^2=14\zeta(3)N_c\alpha_s/\pi$.
When the density of dipoles at a given impact parameter is greater 
than one, multiple scatterings become important.
Thus the scattering probability is not proportional to $n$, but is
an infinite series containing all multiple scattering terms
\cite{kovchegov}.

For our argument, it is only important that 
once the density of dipoles at some impact parameter $r$ becomes 
larger than some fixed critical number, the scattering amplitude 
at this impact parameter
saturates. The exact value of this number depends on the target,
but importantly it does not depend on rapidity. Thus the total
cross section is given by the square of the largest impact
parameter at which the dipole density in the projectile wave
function is of order unity. In order to estimate this directly from
eq. (\ref{density}), we must choose the dipole size $x$
in (\ref{density}) to be the smallest size which is saturated on the
target at initial rapidity. Within the ansatz of eq. (\ref{golec})
this would be $x=Q_s^{-1}(t_0)$. We then find
\begin{eqnarray}
  R^2(t)&=&{1\over 16}{x_0\over Q_s(t_0)}\, 
  \exp\lbrack { \alpha_s N_c\over \pi} \epsilon t\rbrack\, ,
  \label{radius1}\\
  \epsilon&=&7\, \zeta(3) \Big [-1+\sqrt{1+ {8\ln 2 / 7 \zeta(3)}} \Big ]\, .
  \label{epsilon}
\end{eqnarray}
Thus, also for a coulour singlet projectile, we arrive again at the 
exponential growth of the cross section.

The exact value of $\epsilon$ given in eq. (\ref{epsilon})
should not be taken too seriously. The explicit form of the dipole 
density eq. (\ref{density}) was derived by a saddle point integration, 
and as such is valid only for $\ln {16r^2\over x_0 x}<\alpha_st$. 
This condition is not satisfied by eq. (\ref{radius1}). However,
even beyond the saddle
point approximation the density has the form
\begin{equation}
  n(x_0,x,r,t)={1\over x^2}\ln {16r^2\over x_0 x}
  \exp\hspace{-.1cm}
  \Big [ \alpha_s t F({\ln {16r^2\over x_0 x}\over \alpha_s t})
  \Big ]\, .
  \label{density1}
\end{equation}
The relevant condition is $F=0$. Thus, while our calculation
does not specify the numerical value of $\epsilon$, the correct
solution parametrically is the same as (\ref{radius1}).

In the target rest frame, 
this violation of unitarity by the BK evolution can be understood
as follows: Start with a single dipole scattering on the hadronic target
of transverse size $R_{\rm target}$. With increasing energy the projectile 
dipole emits 
additional dipoles strictly according to the BFKL evolution. The density
as well as the transverse size of the projectile state thus grows.
The increase in density leads to increasing importance of multiple
scatterings which are properly accounted for in the BK derivation.
This ensures that the scattering probability saturates locally.
In the saturation regime,
as long as the size of the projectile state $R(t)$ is smaller than the
target size $R_{\rm target}$, the cross section grows essentially
due to surface effects, 
\begin{equation}
  \sigma = \pi R^2_{\rm target}+2\pi R_{\rm target}x_0
  \exp\hspace{-.1cm}\Big [ {\alpha_s N_c\over 2\pi} \epsilon t\Big ]\, .
\end{equation}
Thus as long as $\alpha_s \epsilon t<\ln{R_{\rm target}\over x_0}$,
the cross section is practically geometrical. However once the
energy is high enough so that the projectile size is larger than
that of the target, the total cross section is determined by the former
and grows exponentially with the logarithm of energy according to
eq. (\ref{radius1}).

This also illustrates that the applicability of the BK
evolution crucially depends on the nature of the target.
If the target is thick enough, so that the 
multiple scatterings become important before the growth of the 
projectile radius does,  and if the target is wide enough, 
so that saturation occurs before the
projectile radius swells beyond that of the target, then there
is an intermediate regime in which the inelastic cross section
remains practically constant and equal to $\pi R^2_{\rm target}$.
Then BK applies. However, if the target is a nucleon, neither one 
of these conditions is satisfied. Thus the tainted infrared behaviour
of the BFKL evolution of the projectile will show up right away
and will invalidate the application of the BK equation.

In order to discuss the violation of unitary from the point of view 
of the evolution of target fields, we now go back to the stochastic 
process (\ref{wei}).
The RHS of (\ref{wei}) describes the total Coulomb
(Weizs\"acker-Williams) field at point $x$ due to the colour charge
sources at points $z$. Since the noise is stochastic, the
colour sources are completely uncorrelated both in the transverse
plain and in rapidity. For this random source, the square of the
total colour charge is proportional to the area, and this is
precisely the factor $R^2$ in eq. (\ref{langevinsol}). The incoming
dipole thus scatters on the Coulomb field created by the large
incoherent colour charge. Because the Coulomb field is long range,
the whole bulk of the region populated by the sources contributes
to the evolution and leads to rapid growth of $R$. If the field
created by the sources was screened by some mass, the evolution
would be unitary. To illustrate this point, we
substitute the Coulomb field $(x-z)_i/(x-z)^2$ in (eq.\ref{wei}) by
an exponentially decaying field $m\exp\{-m|x-z|\}$. It is
straightforward to perform now the same analysis as before.
Eq. (\ref{appr}) is replaced by
\begin{equation}
  \int_{|z|<R(\tau)} \hspace{-0.7cm}
  d^2z\, m^2\exp\{-m|x-z|\}=\exp\{-m|x-R|\}\, .
  \label{appr1}
\end{equation}
This leads to the substitution $R^2\rightarrow \exp\{mR\}$
in all subsequent equations with the end result that
\begin{equation}
  R(t)=\alpha_s{\epsilon\over m}t\, ,
\end{equation}
which in fact saturates the Froissart bound.
Thus the reason for the violation of unitarity is that the evolution is
driven by the emission of the long range Coulomb field from a large number
of {\it incoherent} colour sources in the target.

Cutting off the Coulomb field is not the only possibility to cure this
problem. Another option is that the sources of the colour charge in the 
high density regime cease to be incoherent. If they have 
correlations ensuring that the total colour charge in a region of fixed 
size $L$ is zero, then the incoming dipole would feel the Coulomb field 
only within the fixed distance $L$ from the black region. Thus the new 
charges produced by the evolution would only "split off" the edges
of the black region rather than from its bulk. This scenario is equivalent 
to exponential decay of the field, and will lead to a unitary evolution. 
In a confining theory like QCD, it is likely to be materialized. We note 
that the desirability  of such colour charge correlations was stressed 
in a somewhat different context in \cite{lam}.

Although such charge correlations do not arise in the BK evolution,
it is not {\it a priori} clear that they are not present in a more
complete semiperturbative framework which still does not take into
account the physics of confinement at low energies. In fact, the BK 
framework is incomplete inasmuch as it takes the evolution of the 
projectile wave function to be pure BFKL. One expects that once the 
density of gluons in this wave function 
becomes large, interactions should lead to 
saturation effects on the wave function level, i.e., the density of 
the dipoles should grow slower than eq. (\ref{density}). Such corrections
should still be semiperturbative, in the sense that they are present 
at small $\alpha_s$. However, for scattering on "small" targets, they
will become important at the same energy as the multiple scattering 
terms resummed in eq. (\ref{bal},\ref{kov}). 
These wave function saturation effects 
% (or pomeron loops in the language of \cite{kovchegov}) 
may lead to charge correlations 
of the type necessary to unitarize the total cross section.

We finally note that the unitarity bound for Deeply Inelastic 
Scattering is different. In this case, the projectile is a virtual 
photon without fixed hadronic size. For transverse polarization, 
its perturbatively known wave function is 
$\Phi^2(r)\propto \alpha_{em}{1\over r^2}$
for $r^2 \ll Q^{-2}$. 
In such a projectile not all dipoles saturate at the same energy.
The main contribution to the scattering probability comes from
dipoles of size $r > Q_s^{-1}(t)$  
which are saturated.
At high energy (i.e. for $Q_s^{-1}\ll Q^{-1}$),
\begin{equation}
N(\gamma^*)=\int_{Q_s^{-2}<r^2<Q^{-2}} d^2r
\Phi^2(r)\propto\alpha_{em}\ln{Q_s/Q}\, .
\end{equation}
With the exponential dependence (\ref{qexp}) of $Q_s$ on rapidity this
translates into
$  N(\gamma^*)\propto\alpha_{em}\alpha_s \ln s/s_0$, 
and therefore
\begin{equation}
\sigma_{\rm DIS}\propto\alpha_{em}\alpha_s\pi R^2(t)t\, .
\end{equation}
The DIS cross section has an extra power of $t$ relative to
the cross section of a purely hadronic process. This extra power
of $t$ is consistent with the numerical results of \cite{gms}.

{\bf Acknowledgements} This work has been supported in part by
PPARC. We thank G. Milhano and H. Weigert for helpful discussions.
U.A.W. thanks the Department of Mathematics and Statistics,
University of Plymouth for hospitality while part of this work was
done.

\end{document}